\documentclass[aps,prx,twocolumn,amsmath,amssymb,superscriptaddress]{revtex4-1}
\usepackage{graphicx}
\usepackage{dcolumn}
\usepackage{graphics}
\usepackage{bbm}
\usepackage{amsmath,amsfonts,amssymb}
\usepackage{wrapfig}
\usepackage[colorlinks,linkcolor=black, citecolor=black,urlcolor=black,bookmarks=false,hypertexnames=true]{hyperref} 
\usepackage{color}

\begin{document}
\newcommand{\dggedits}[1]{\textcolor{blue}{#1}}

\title{Zero-field Edge Magnetoplasmons in a Magnetic Topological Insulator}

\author{A. C. Mahoney}
	\thanks {These authors contributed equally to this work.}
	\affiliation{ARC Centre of Excellence for Engineered Quantum Systems, School of Physics, The University of Sydney, Sydney, NSW 2006, Australia.}
\author{J. I. Colless}
	\thanks {These authors contributed equally to this work.}
	\affiliation{ARC Centre of Excellence for Engineered Quantum Systems, School of Physics, The University of Sydney, Sydney, NSW 2006, Australia.}

\author{L. Peeters}
	\thanks {These authors contributed equally to this work.}
	\affiliation{Department of Physics, Stanford University, Stanford, California 94305, USA.}
\author{S. J. Pauka}
\affiliation{ARC Centre of Excellence for Engineered Quantum Systems, School of Physics, The University of Sydney, Sydney, NSW 2006, Australia.}
\author{E. J. Fox}
\affiliation{Department of Physics, Stanford University, Stanford, California 94305, USA.}
\affiliation{Stanford Institute for Materials and Energy Sciences, SLAC National Accelerator Laboratory, 2575 Sand Hill Road, Menlo Park, California 94025, USA. }
\author{X. Kou} 
\affiliation{Department of Electrical Engineering, University of California, Los Angeles, California 90095, USA.}
\author{Lei Pan}
\affiliation{Department of Electrical Engineering, University of California, Los Angeles, California 90095, USA.}
\author{K. L. Wang}
\affiliation{Department of Electrical Engineering, University of California, Los Angeles, California 90095, USA.}
\author{D. Goldhaber-Gordon}
\email[Corresponding author: ] {goldhaber-gordon@stanford.edu}
\affiliation{Department of Physics, Stanford University, Stanford, California 94305, USA.}
\affiliation{Stanford Institute for Materials and Energy Sciences, SLAC National Accelerator Laboratory, 2575 Sand Hill Road, Menlo Park, California 94025, USA. }
\author{D. J. Reilly}
\email[Corresponding author: ] {david.reilly@sydney.edu.au}
\affiliation{ARC Centre of Excellence for Engineered Quantum Systems, School of Physics, The University of Sydney, Sydney, NSW 2006, Australia.}
\affiliation{Microsoft Station Q Sydney, Sydney, NSW, 2006, Australia.}
\maketitle

{\bf{
Incorporating ferromagnetic dopants, such as chromium \cite{chang2013experimental, kou2014scale, checkelsky2014trajectory} or vanadium \cite{chang2015high}, into thin films of the three-dimensional (3D) topological insulator (TI) (Bi,Sb)$_2$Te$_3$ has recently led to the realisation of the quantum anomalous Hall effect (QAHE), a unique phase of quantum matter. These materials are of great interest \cite{haldane1988model,hasan2010colloquium},
since they may support electrical currents that flow without resistance via edge channels, even at zero magnetic field. To date, the QAHE has been investigated using low-frequency transport measurements \cite{chang2013experimental,kou2014scale, checkelsky2014trajectory, bestwick2015precise, chang2015high}. However, transport requires contacting the sample and results can be difficult to interpret due to the presence of parallel conductive paths, via either the bulk or surface \cite{chang2013experimental, bestwick2015precise}, or because additional non-chiral edge channels may exist \cite{wang2013anomalous, kou2014scale, PhysRevLett.115.057206}. Here, we move beyond transport measurements by probing the microwave response of a magnetised disk of Cr-(Bi,Sb)$_2$Te$_3$. We identify features associated with chiral edge-magnetoplasmons (EMPs), a signature that robust edge-channels are indeed intrinsic to this material system. Our results provide a measure of the velocity of edge excitations without contacting the sample, and pave the way for a new, on-chip circuit element of practical importance: the TI, zero-field microwave circulator. }}

It is now understood that ferromagnetism, by lifting spin degeneracy and breaking time reversal symmetry at zero magnetic field, can transform a topological insulator (TI) into a new phase of matter that hosts chiral edge states \cite{qi2008topological, liu2008quantum, yu2010quantized, qi2011topological, nomura2011surface}. The signature of this phase is the quantum anomalous Hall effect (QAHE), in which the transverse conductance of a magnetised Hall bar remains quantised in units of the conductance quantum, even in the absence of an external magnetic field. Given that bulk insulators and ferromagnets are commonplace at room temperature, there is optimism that the QAHE may not be limited to the cryogenic regimes of today's experiments. A room-temperature QAHE in which edge states propagate without dissipation could impact some of the challenges facing current-generation high speed integrated circuits. The presence of robust edge states in these material systems also opens the prospect that they support EMP excitations, resonant drum-modes of the electron gas that are well-known in the context of the quantum Hall effect \cite{volkov1988edge, ashoori1992edge,talyanskii1992spectroscopy}. These resonant modes typically occur at microwave frequencies, and are distinct from the predicted plasmonic phenomena of TI materials near optical frequencies \cite{Rudner_PNAS}. Beyond their fundamental interest, the velocity of EMP excitations is typically reduced compared to the speed of light, making them ideal platforms for constructing on-chip delay networks, high-impedance transmission lines, and non-reciprocal devices such as gyrators and circulators needed for quantum information processing \cite{viola2014hall, mahoney2016chip, bosco2016self}.  

Here we investigate the zero-field magneto-plasmonic response of a magnetised, contactless disk of the ferromagnetic TI Cr-(Bi,Sb)$_2$Te$_3$.  The fabrication of both Hall bars and resonant disk structures enables us to make a one-to-one comparison between transport data and the microwave excitation spectrum of the material. By implementing a three-port circulator configuration, we show that the low frequency plasmon response exhibits non-reciprocal behaviour, consistent with chiral EMPs.  The existence of EMPs in the disk and their correspondence with a minimum in the longitudinal resistance of the Hall bar provide further convincing evidence that this system supports a robust edge state. Finally, we examine the dependence of circulation on excitation power and temperature, suggesting that microwave measurements can serve as a sensitive probe of the conditions at the edge. 

\begin{figure*}\centering
	\includegraphics[scale=0.38]{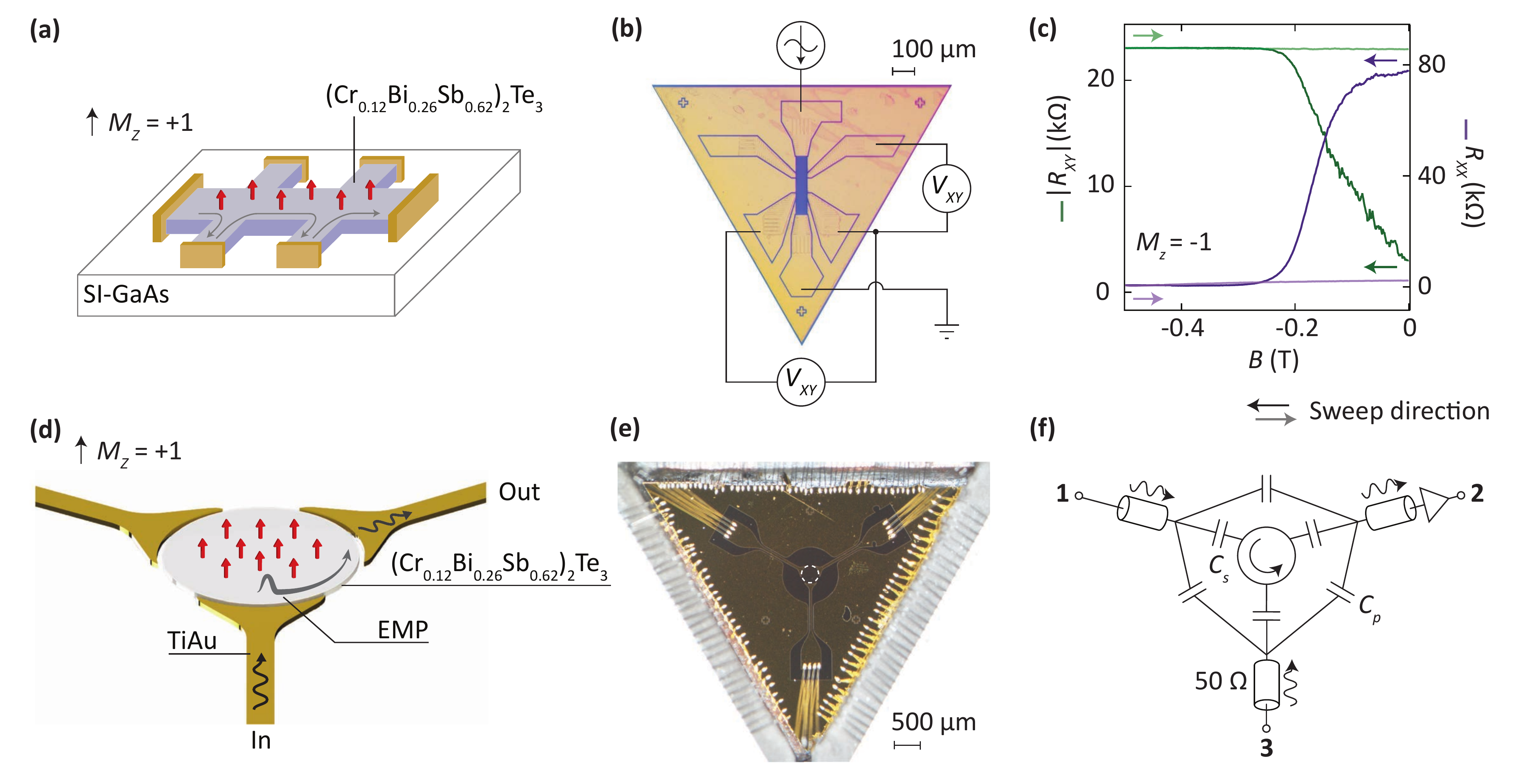}
	\caption{\label{fig:photos}\textbf{Experimental set-up.} \textbf{(a)} Illustration of the quantum anomalous Hall effect in a three-dimensional topological insulator thin film with ferromagnetic dopants. \textbf{(b)} Photograph of a Hall bar fabricated on a device with seven quintuple layers of (Cr$_{0.12}$Bi$_{0.26}$Sb$_{0.62}$)$_2$Te$_3$ grown epitaxially on a GaAs substrate. Standard lock-in techniques are used to measure the transport properties of the material. \textbf{(c)} Transverse and longitudinal resistance ($R_{xy}$ and $R_{xx}$) as the perpendicular magnetic field is swept out to -0.5 T (dark coloured lines), and then back to zero field (light shaded lines). \textbf{(d)} Cartoon of a three-port circulator device with a magnetic topological insulator.  \textbf{(e)} Photograph of the circulator device. \textbf{(f)} A circuit schematic for the experimental setup. The parasitic capacitances between port electrodes ($C_p$) and from the ports to the EMP modes ($C_s$) are indicated. Port 2 is connected to a low-noise cryo-amp operating at 4 K, allowing measurement of $S_{21}$ and $S_{23}$ through a common output line.}
	\vspace{-0.5cm}
\end{figure*} 

Turning to the experimental setup shown in Fig.~1, the magnetic three-dimensional (3D) TI used to make the circulator and corresponding Hall bar is seven quintuple layers of (Cr$_{0.12}$Bi$_{0.26}$Sb$_{0.62}$)$_2$Te$_3$. The film is grown on a semi-insulating (111)B GaAs substrate by molecular beam epitaxy, then capped with alumina to protect the surface. To define the microwave circulator, we use photolithography to pattern a circular, 330 $\mu$m diameter mesa and etch away the remaining film via Ar ion milling (see supplemental material). We next pattern capacitive contacts and a ground plane, depositing 120 nm Au with a Ti sticking layer by e-beam evaporation. The contacts are designed to be 20 $\mu$m away from the mesa edge.

Starting at zero field, the transport data in Fig.~1(c) show the longitudinal and transverse resistances of the Hall bar, $R_{xx}$ and $R_{xy}$, during the initial magnetisation sequence, sweeping the field from zero to -0.5 T at the cryostat base temperature of $T$ = 20 mK (dark purple and green lines). As the field is applied for the first time we observe $R_{xx}$ drop from $\sim$ 80 k$\Omega$ to $\sim$ 500 $\Omega$ as $R_{xy}$ increases towards the resistance quantum = $h/e^2$ ($h$ is Planck's constant and $e$ the electron charge). After waiting several hours the field is swept back to zero, with transport data in this direction shown as lightly-shaded lines in Fig.~1(c). Following this initial magnetisation sequence, we observe the signatures of the QAHE, namely that $R_{xx}$ remains near zero and $R_{xy}$ does not vary by more than 1$\%$. An accurate and precise measurement of $R_{xy}$ requires accounting for possible geometric effects of the contacts and calibration using a known resistance standard, as was done in Ref.~\cite{bestwick2015precise}. In the absence of these corrections, the measured resistance plateau value of 25750 $\Omega$ is within the uncertainty expected for the quantum of resistance, 25813 $\Omega$ \cite{chang2013experimental, kou2014scale, checkelsky2014trajectory}.

\begin{figure}\centering
	\includegraphics[width=\columnwidth]{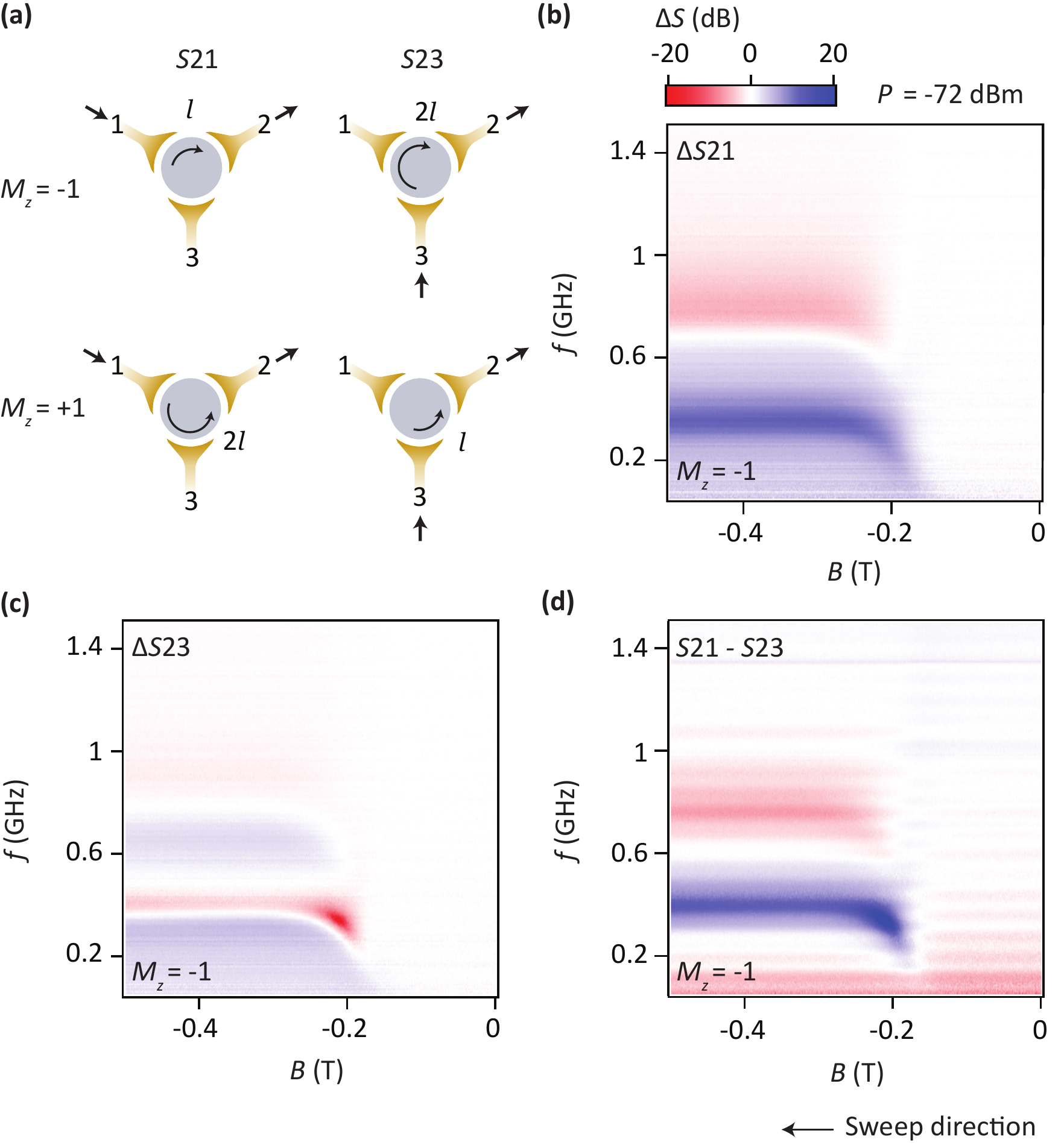}
	\caption{\textbf{Initial magnetisation response.} \textbf{(a)} Illustration of chiral edge transport in a circulator setup for different magnetisation and port configurations. Arc path lengths are denoted as $l$ and 2$l$. \textbf{(b)} and \textbf{(c)} show the microwave response of signals excited from ports 1 and 3 and amplified out of port 2, as the  magnetic field is varied. These traces have been normalised to the reciprocal background prior to sample magnetisation. Isolation $S_{21} - S_{23}$ is shown in \textbf{(d)}, where the difference between the bare S-parameter traces is plotted without background normalisation. Past the coercive field of -0.16 T, strong non-reciprocity is observed at two distinct frequency bands, with opposite amplitudes.}
	\vspace{-0.5cm}
	\label{fig2}
\end{figure}

\begin{figure*}\centering
	\includegraphics[scale=0.55]{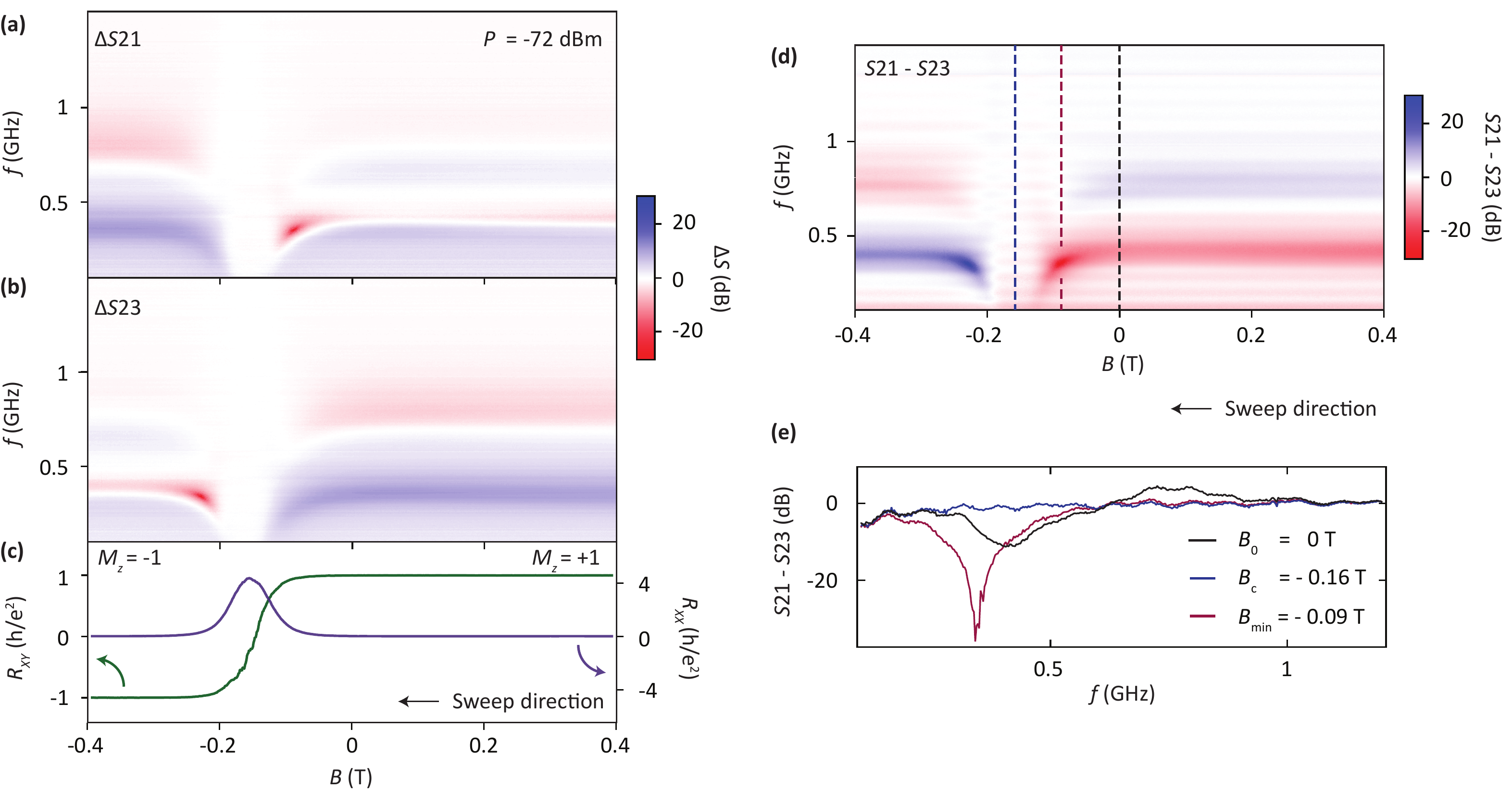}
	\caption{\textbf{Microwave isolation across positive and negative magnetisation directions.}  \textbf{(a, b)} Normalised $\Delta S_{21}$ and $\Delta S_{23}$ responses measured at the same time as transport data in \textbf{(c)}. A pre-magnetisation frequency-dependent background has been subtracted from each of the traces in \textbf{(a,b)}. 
	 \textbf{(c)} Transverse (green) and longitudinal (purple) dc resistances measured on a Hall bar as the magnetisation direction is swept from positive to negative. Away from the coercive field, raw $R_{xy}$ approaches the resistance quantum while $R_{xx}$ measures $\sim$ 500 $\Omega$. In \textbf{(d)} we compare the difference between the bare $S_{21}$ and $S_{23}$ paths, providing a measure of isolation in the system. \textbf{(e)} Shows cuts through \textbf{(d)} at zero applied magnetic field ($B_0$, black line), the coercive field ($B_{\textrm c}$, blue) and at the point where a power minimum is observed ($B_{\textrm min}$, mauve).}
	\vspace{-0.5cm}
\end{figure*}

\begin{figure}
	\includegraphics[scale=0.4]{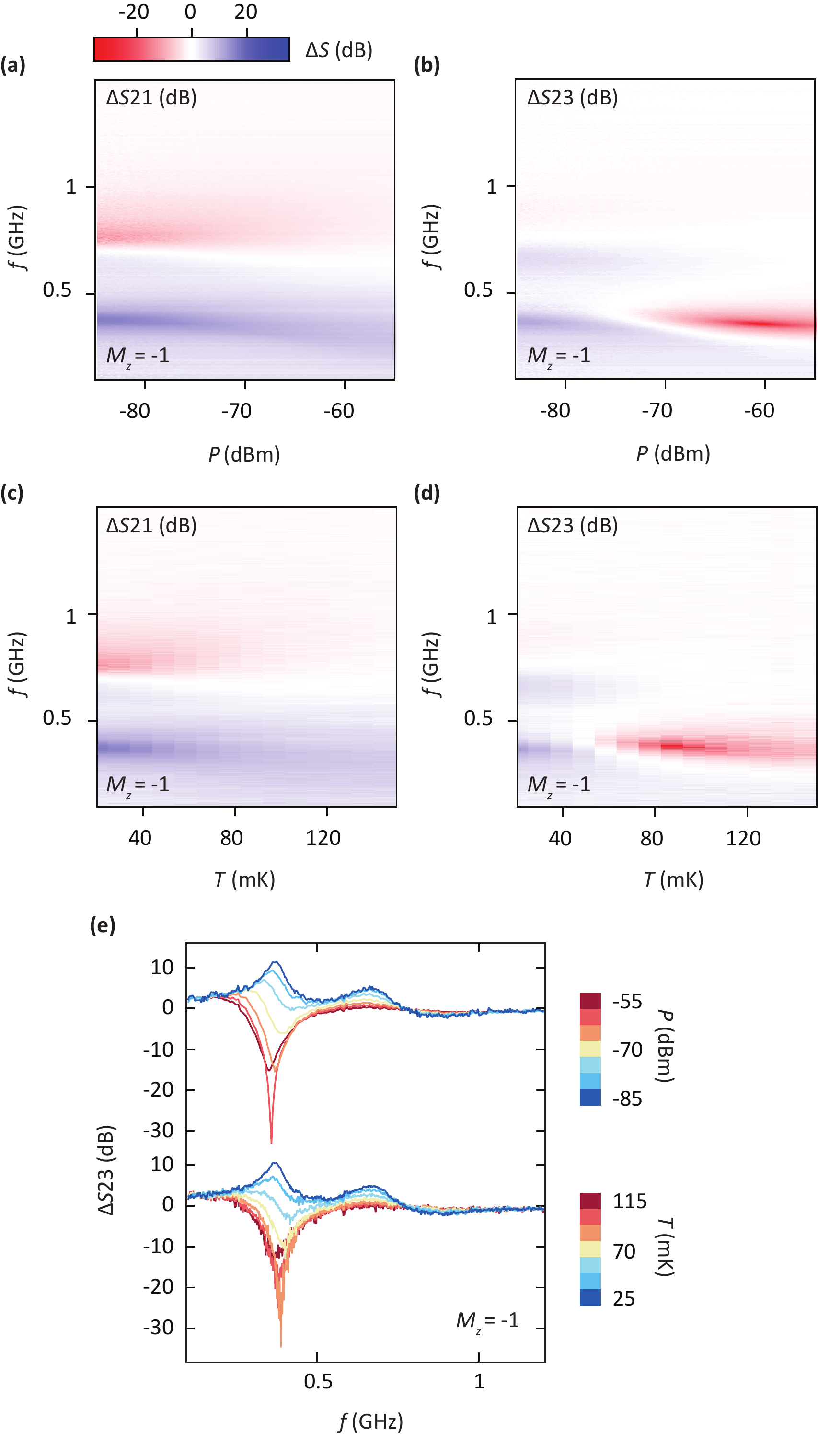}
	\caption{\textbf{Effect of temperature and microwave power.} \textbf{(a,b)} Show the frequency response $\Delta S_{21}$ and $\Delta S_{23}$ with increasing microwave power applied to the excitation port. For these measurements $T \sim$ 20 mK. In \textbf{(c)} and \textbf{(d)} the dependence of $\Delta S_{21}$ and $\Delta S_{23}$ is measured as a function of cryostat temperature while the applied port power is kept low at -87 dBm. In \textbf{(a-e)}  all data are taken at zero applied magnetic field after the sample has been magnetised with $M_z=-1$. Akin to Figures 2 $\&$ 3, the same pre-magnetisation background taken at constant power and cryostat base temperature is subtracted from each of the 2D plots. \textbf{(e)} Presents 1D cuts through the colour plots of $\Delta S_{23}$ in \textbf{(b)} and \textbf{(d)}. The top (bottom) panel depicts traces at constant power (temperature), showing the evolution of the Fano-like lineshape.}
	\vspace{-0.5cm}
\end{figure}
We compare these transport data with the microwave response of an etched TI disk on the same material, configured as a circulator as shown in Fig.~1(d-f). Similar designs comprising an rf excitation and detection port have been used to probe EMPs in the quantum Hall effect regime in GaAs semiconductors \cite{kamata2010voltage,kumada2011edge, mahoney2016chip} and in graphene \cite{petkovic2013carrier, kumada2013plasmon, kumada2014resonant}. These EMP modes are charge density waves supported by edge channels at the boundary of the material. For traditional semiconductor samples, such as GaAs, the propagation velocity and therefore microwave frequency response of the EMP is set by the ratio of the electric and magnetic field at the edge. To what extent this picture holds true for TI materials remains an open question motivating our work.

In our setup, a 3-port design further allows the non-reciprocal character of the device to be probed by determining whether the signal traverses the TI disk via a left-handed or right-handed path \cite{viola2014hall, mahoney2016chip}. The experiment comprises $S$-parameter measurements in which the amount of microwave power transmitted from port 1 to port 2 ($S_{21}$), or port 3 to port 2 ($S_{23}$) is detected as a function of external magnetic field and magnetisation state ($M_z = M/\left| M\right|= \pm 1$) of the TI disk (see Fig.~2(a)). The two symmetrical paths, port 1 to 2 and port 3 to 2, are designed to be equivalent in the absence of chiral transport.

We note that for any measurement, microwave power can be coupled directly and reciprocally between the ports via the geometric (parasitic) capacitance that shunts the disk ($C_p$ in Fig.~1(f)). In our sample geometry $C_p$ is estimated to be in the vicinity of a few hundred femto-Farads \cite{mahoney2016chip}. The combination of a direct capacitive path in parallel with the conductive edge-channels in the disk creates an interferometer in which signals travel at the speed of light in the capacitive arm, and at a velocity in the other arm that is determined by the magnetoplasmonic response \cite{mahoney2016chip}. It is primarily the difference in velocities (and, to a lesser extent, path lengths)
between the two paths of the interferometer that yields a phase offset between the two signals when they recombine at the receiving port. Further, the amplitude of the signals is set by the impedances of the two paths; if the EMP resonator has a moderate Q-factor, these amplitudes can easily be made comparable by driving the circuit at a frequency slightly detuned from the resonant mode of the EMP. Within this interference picture, the response of the circuit can be interpreted as a Fano resonance that depends on the length of path traveled by the EMP, $l$ or 2$l$ depending on the excitation port and magnetisation direction (see Fig.~2(a)). With port 2 always set to be the receiving port, transmitting power from port 1 or port 3 thus configures the edge length to be $l$ or 2$l$. The difference between the two paths' transmission, measured as $S_{21} - S_{23}$, determines the isolation or non-reciprocity of the circuit. 
 
We begin the presentation of the circulator data in a similar fashion to the transport measurements, starting at zero magnetic field and recording $S_{21}$ and $S_{23}$ as the external field is stepped out to -0.5 T. To enable a direct comparison between transport and the microwave response of the disk during the one-off magnetisation sequence, we acquire transport data as well as $S_{21}$ and $S_{23}$ at a fixed magnetic field before stepping the field (i.e., all data in Fig.~1(c) and Fig.~2 were obtained concurrently). All individual $S_{21}$ and $S_{23}$ data throughout the paper are normalised (denoted by $\Delta$) by subtracting the initial frequency dependence of the signal in logarithmic units (dB) in the unmagnetised state, $B=0$ and $M_z=0$, where the response is reciprocal. This calibration trace is taken at cryostat base temperature ($T \sim$ 20 mK) and a port excitation power of -72 dBm.  This normalisation alleviates frequency-dependent artefacts, for instance transmission oscillations due to line impedance mismatch that do not evolve with magnetic field (see supplemental material). 

Forward transmission $\Delta S$ is shown in Fig.~2(b) and 2(c) as a function of frequency and magnetic field for the two paths $S_{21}$ and $S_{23}$. As the sample begins to magnetise at the coercive field ($\sim$ -0.16 T), we observe resonance-like dips and peaks in the frequency spectrum of the disk, evident as red and blue coloured horizontal bands appearing at the field strength where $R_{xy}$ (Fig.~1(c)) approaches the resistance plateau $h/e^2$. This is the microwave signature of the QAHE. Compared with EMPs in 2 dimensional electron systems \cite{mahoney2016chip} where the frequency $\omega_{EMP}$ is proportional to $1/B$, in the TI we observe a flat dispersion as a function of magnetic field, centred at  the fundamental mode of the EMP, near 400 MHz. This is consistent with dc transport measurements of the Hall resistance which takes on a constant, quantised value after the sample is magnetised. Measuring the frequency at which these resonances occur in combination with the circumference of the TI disk gives an EMP velocity at the fundamental mode of $\sim 4 \times 10^5$ m/s, similar to what is found in other structures comprising stacks of semiconductors \cite{kamata2010voltage, kumada2011edge, mahoney2016chip}. 

The microwave response shows that  the parameters $S_{21}$ and $S_{23}$ deviate from each other as the disk becomes magnetised. This is a result of the non-reciprocity of the system, evident in Fig.~2(d) where we have subtracted the bare $S$-parameters ($S_{21} - S_{23}$) from each other to show the difference between the two configurations of the circulator. Again, we interpret these measurements of $S_{21}$ and $S_{23}$ as characterizing paths around the edge of the disk in the same (chiral) direction with arc length $l$ and 2$l$ (Fig.~2(a)). Considering the measurement in Fig.~2(d), it is apparent that microwave power can both circulate near the fundamental EMP frequency (blue frequency band) and `counter circulate' in an opposite direction near the first harmonic (red frequency band). This behaviour is also observed for GaAs devices in the quantum Hall regime \cite{mahoney2016chip} and is understood to arise from a Fano-like interference between the slow-velocity resonantly circulating edge mode and the less frequency-sensitive parallel capacitive path \cite{placke2016model}. We remark that the observation of circulation and counter-circulation is a further signature of the plasmonic response of the chiral edge state. 

The quantum anomolous Hall effect is unique in that it supports a robust chiral edge state at zero applied field \cite{bestwick2015precise}. To examine the zero-field response of the magnetised TI system, we continue to take transport measurements on the Hall-bar concurrent with $S$-parameter data on the circulator, as the system is swept from positive to negative field through zero, as shown in Fig 3(a) and 3(b). The transport data in Fig.~3(c) show the familiar signature of the QAHE with $R_{xx}$ peaking and $R_{xy}$ switching sign at the coercive field indicated by the blue dashed line ($\sim$ -0.16 T). At $B$ = 0, the system remains magnetised with $R_{xy}$ reaching a maximum value of 25.75 k$\Omega$. 

In comparison to the transport measurements, the microwave response of the TI reveals new information. The response of the disk for each of the signal configurations, characterised by $\Delta S_{21}$ in the top panel and $\Delta S_{23}$ in the middle panel, is strongly asymmetric about the coercive field. Symmetry is restored, however, if in addition to the sign of the magnetic field the ports are also interchanged, so that the red band in $\Delta S_{21}$ on the left of the coercive field mirrors the red band in $\Delta S_{23}$ on the right, and vice versa for blue features. This strong non-reciprocity is most evident in the differential form of the data $S_{21} -  S_{23}$, as shown in Fig.~3(d). At zero field the circulator continues to exhibit non-reciprocity $\sim$ 10 dB (Fig.~3(e)). Intriguingly, the device is maximally non-reciprocal at a field approaching the coercive field, producing a `hot-spot' feature in the $S$-parameter response (mauve dashed-line Fig.~3d). As described below, we suggest these features are linked to enhanced dissipation. 

Finally, we investigate the temperature and microwave power dependence of the EMP spectra in an effort to better understand the details of the zero-field edge state.  At $B=0$ and with the TI magnetised ($M_z=-1$), $\Delta S_{21}$ and $\Delta S_{23}$ are measured as a function of applied microwave power $P$, as shown in Fig.~4(a) and 4(b). In addition to the usual non-reciprocity at constant power, we observe an evolution in the non-reciprocal features with increasing $P$ that is dependent on the length of the edge segment. While the response of $\Delta S_{21}$ (characterized by arc length $l$) begins to fade out at high powers, the amplitude of $\Delta S_{23}$ (2$l$) changes sign near the fundamental frequency and exhibits a pronounced minimum or hot-spot near $P$ = -60 dBm. Interchanging the ports and repeating the measurement at $M_z=+1$ and $B=0$ produces similar features in accordance with a reversal of chirality (see supplemental material). Mirroring the dependence with power, increasing the cryostat temperature also produces a change of sign relative to the pre-magnetised state for the longer edge path (2$l$): Raising $T$ as in Fig.~4(c) and (d) leads to $\Delta S_{21}$ becoming gradually washed out, while $\Delta S_{23}$ produces a hot-spot around $T$ = 85 mK. This effect is further illustrated in Fig.~4 (e), where 1D cuts at constant power (top panel) or temperature (bottom panel) are shown for $\Delta S_{23}$.

So hot-spots -- strong decrease in the microwave power transmitted between ports -- occur at particular magnetic fields, powers, or temperatures. Appealing to the phenomenology of the interferometer pictured in Fig.~1(f), we note that the direction of power transmission between ports, either circulation or counter-circulation, is set by the relative phase and amplitude of the signals in the edge-state arm compared to the direct capacitive arm of the interferometer. In general, this picture accounts for the constructive interference of signals for the shorter edge path ($l$) and destructive interference for the longer (2$l$), when driving near resonance of the EMP fundamental. Extending this picture to include dissipation, we suggest that losses in either arm of the interferometer can adjust the relative amplitudes. As noted above, the relative phase of the two paths is naturally tuned by $\pi$ upon crossing the EMP resonance. At a frequency where the relative phase is $\pi$ we might expect a sign change in the relevant $S$ parameter response. In this way the hot-spots can be understood as regions where almost perfect cancellation between the two arms occurs. The appearance of dissipation in either path with increasing temperature, microwave power, or near the coercive field is perhaps not unexpected. Accounting for the microscopic mechanisms underlying such dissipation remains an open challenge \cite{PhysRevLett.115.057206, wang2013anomalous, bestwick2015precise,wang2014universal, kou2015metal, feng2015observation, li2016origin}, given that all measurements are well below both the Curie temperature of these ferromagnetic TIs (of order 10 to 20 K) and the energy scale of the exchange gap as measured by ARPES (of order 10 meV). 

To conclude, we have probed the EMP spectrum of a magnetic topological insulator, comparing its microwave response to transport data. The measurement setup can be understood as an interferometer with the disk of TI in one arm of the interferometer and a parasitic capacitance in the other. Within this picture, the response of the system exhibits resonances that can be explained by accounting for the slow velocity of EMPs as they traverse an arc-length of the TI disk's edge rather than the bulk. In addition to the device examined, we have examined a second circulator, fabricated from a separately-grown wafer, and found it to exhibit strikingly similar behaviour in all aspects (see supplemental material). We suggest that this similarity between devices is related to the robust properties of the edge state, set by the non-trivial topology of the material system rather than, for instance, the specific configuration of microscopic disorder. Taken together, our microwave measurements provide strong evidence that this material system indeed supports robust, chiral edge states at zero magnetic field, opening the prospect of compact microwave components based on magnetic topological insulators. 
\\ 

\noindent {\bf{Acknowledgements}}\\
We thank Andrew Doherty and David DiVincenzo for useful conversations. Device fabrication and preliminary characterization was supported by the Department of Energy, Office of Science, Basic Energy Sciences, Materials Sciences and Engineering Division, under Contract DE-AC02-76SF00515. Microwave and transport measurements presented in the main text were supported by Microsoft Research, the Army Research Office grant W911NF-14-1-0097, the Australian Research Council Centre of Excellence Scheme (Grant No. EQuS CE110001013). Materials growth was supported by the DARPA MESO program under Contracts No.\ N66001-12-1-4034 and No.\ N66001-11-1-4105. Infrastructure and cryostat support were funded in part by the Gordon and Betty Moore Foundation through Grant GBMF3429. E.~J.~F. acknowledges support from a DOE Office of Science Graduate Fellowship. L.~P. was supported by a Stanford Graduate Fellowship.  \\

\noindent {\bf{Competing financial interests}}
\newline The authors declare no competing financial interests.
\newline


\end{document}